\documentclass[reprint, amsmath,amssymb, aps, superscriptaddress]{revtex4-2}
\usepackage{amsmath,amsfonts,amssymb,amsthm,graphics}
\usepackage[next]{inputenc}
\usepackage[colorlinks=true,citecolor=blue,linkcolor=blue]{hyperref}
\usepackage{bbm}
\usepackage{bbold}
\usepackage{booktabs}
\usepackage{multirow}
\usepackage{hhline}
\usepackage{comment}
\usepackage{float}
\usepackage{latexsym}
\usepackage[dvipsnames]{xcolor}
\usepackage{graphicx}
\usepackage{bm}
\usepackage{verbatim}
\usepackage{siunitx}
\usepackage{physics}
\usepackage{subfigure}

\newcommand{\PT}{$\mathcal{P}\mathcal{T}$}
\renewcommand{\vec}{\mathbf}

\begin{document}

\title{Oscillatory Edge Modes in Two Dimensional Spin-Torque Oscillator Arrays}
\author{Shivam Kamboj}
\affiliation{Department of Physics and Astronomy, San Jos\'{e} State University, San Jos\'{e}, California, 95192, USA}
\affiliation{Department of Physics, University of Califonia, Merced, California, 95343, USA}

\author{Rembert A. Duine}
\affiliation{Institute for Theoretical Physics and Center for Extreme Matter and Emergent Phenomena, Utrecht University, Leuvenlaan 4, 3584 CE Utrecht, The Netherlands}
\affiliation{Department of Applied Physics, Eindhoven University of Technology, P.O. Box 513, 5600 MB Eindhoven, The Netherlands}

\author{Benedetta Flebus}
\affiliation{Department of Physics, Boston College, 140 Commonwealth Avenue, Chestnut Hill, Massachusetts 02467, USA}

\author{Hilary M. Hurst}
\affiliation{Department of Physics and Astronomy, San Jos\'{e} State University, San Jos\'{e}, California, 95192, USA}

\begin{abstract}
Spin torque oscillators (STOs) are dissipative magnetic systems that provide a natural platform for exploring non-Hermitian phenomena. We theoretically study a two-dimensional (2d) array of STOs and show that its dynamics can be mapped to a 2d, non-Hermitian Su-Schrieffer-Heeger (SSH) model. We calculate the energy spectrum and identify the one-dimensional (1d) edge states of our model, corresponding to auto-oscillation of STOs on the boundary of the system while the bulk oscillators do not activate. We show that tuning the Gilbert damping, injected spin current, and coupling between STOs allows for exploring the edge state properties under different parameter regimes. Furthermore, this system admits 1d edge states with non-uniform probability density, and we explore their properties in systems of different sizes. Additional symmetry analysis indicates that these states are not topologically protected but are nevertheless confined to the edge of the system, as the bulk is protected by \PT-symmetry. These results indicate that 2d arrays of STOs may be useful to explore novel edge state behavior in dissipative systems.    
\end{abstract}

\maketitle 

\section{Introduction}

Topology and its connection to condensed matter systems has been the subject of intense research for nearly fifty years, since the discovery of quantized Hall resistance and its topological origin~\cite{Klitzing1980, Thouless1982}. Topology is now understood as a critical underlying feature of many materials that can affect global transport properties, resulting in, e.g., the quantum anomalous and spin Hall effects~\cite{ Maciejko2011, Liu2016} and leading to entirely new classes of topological materials~\cite{Moore2010, Sato2017, Yan2017}. The effects of topology in systems with dissipation, i.e. non-Hermitian systems, can be markedly different from their Hermitian counterparts~\cite{Hu2011, Yao2018, Lieu2018, Gong2018, kawabata2019, Bergholtz2021, ding2022non}. Non-Hermitian systems can exhibit exceptional points~\cite{heiss2012physics, Doppler2016, Liu2019, deng2023exceptional}, i.e. the coalescence of two or more eigenvectors, as well as the non-Hermitian skin effect, a phenomenon where bulk eigenstates localize on the edge of the system~\cite{yokomizo2021scaling, deng2022non, zhang2022universal}. The edge states in non-Hermitian systems can also exist despite the breakdown of the bulk-boundary correspondence~\cite{Yao2018, Lieu2018} and exhibit lasing behavior~\cite{Feng2017, Ota2020}. While several experimental realizations of non-Hermitian phenomena have been observed in photonic~\cite{Feng2017}, acoustic~\cite{Zhu2014}, and electronic circuits~\cite{helbig2020generalized, kotwal2021active}, their experimental exploration in magnonic systems is still in its infancy. Magnonic systems are, however, a natural platform in which to realize non-Hermitian physics because they are always coupled to a surrounding environment and exhibit lossy dynamics~\cite{Hurst2022, Yuan2022}.

Spin torque oscillators (STOs) have recently emerged as a promising platform for harboring non-Hermitian phenomena~\cite{Flebus2020, Hurst2022, wittrock2023nonhermiticity}. These magnetic nanopillars are nanometer sized devices that conduct spin currents via spin transfer torque~\cite{slavin2009}. STOs are dissipative systems because, like all magnetic systems, they are subject to ubiquitous spin non-conserving interactions parameterized by Gilbert damping~\cite{Gilbert1955}. The magnetic dynamics of 1d STO arrays was successfully mapped to a non-Hermitian Su-Schrieffer-Heeger (SSH) model with topologically protected lasing edge states in Ref.~\cite{Flebus2020}. In a subsequent numerical study these edge states were shown to be robust in the presence of additional terms such as dipolar interactions and nonlinear STO behavior~\cite{Gunnink2022}. Here, we examine whether these lasing edge states can be realized in higher-dimensional arrays of STOs.

The 1d SSH model and it's non-Hermitian variants are widely used in condensed matter physics to study topological systems~\cite{Hasan2010, Lieu2018}; 2d non-Hermitian SSH models have been studied in a variety of contexts, where a number of interesting properties such as in-gap topological states and non-trivial bulk-band topology have been found~\cite{Obana2019, liu2019topologically, yuce2019topological, helbig2020generalized, kotwal2021active}. However, most of these models lack a clear experimental implementation, with topological circuits being a notable exception~\cite{helbig2020generalized, kotwal2021active}. The sources of non-Hermitian terms (i.e. energy non-conserving terms) can be difficult to quantify and characterize in experimental platforms. Here, we propose STO arrays as a platform for experimental realization of a non-Hermitian 2d SSH model. Specifically, we focus on a geometry consisting of several 1d chains that are weakly coupled to form a 2d STO array. By introducing this new type of `vertical coupling', we derive a non-Hermitian 2d SSH model that exhibits 1d lasing edge states. 

The manuscript is organized as follows: In Section~\ref{Sec:Model} we map the linearized Landau-Lifshitz-Gilbert (LLG) equation for the magnetization dynamics into a non-Hermitian tight-binding Hamiltonian. In Sec.~\ref{Sec:Results} we discuss the properties of our model including the oscillatory edge states and the symmetry properties of the Bloch Hamiltonian. Finally, we conclude and discuss directions for future work in Sec.~\ref{Sec:Conclusion}.

\section{Model \label{Sec:Model}}

We consider a 2d array of $M \times 2N$ STOs which could be fabricated from individual nanopillars~\cite{Flebus2020} or a multilayer structure~\cite{Gunnink2022}. Here, $M$ is the total number of rows in the array and $N$ indicates the number of unit cells per row, where there are two STOs per unit cell. A single STO consists of a layer of fixed magnetic polarization and a `free' magnetic layer without fixed polarization, separated by a thin metallic spacer. By injecting spin current into the free layer, the fixed layer is driven to precess about its equilibrium direction, which is set by an external applied magnetic field. The dynamics of an isolated STO subjected to a magnetic field $\vec{H}_0 = H_0\hat{\vec{z}}$ and spin current $\vec{J}^{S}_{\eta} = J^{S}_{\eta}\hat{\vec{z}}$ are described by the LLG equation for the magnetization vector $\vec{m}_{\eta, ij}$~\cite{slavin2009}. Here, the index $\eta = A, B$ denotes the $A$ and $B$ sublattices and the indices $i,j$ label the sites using the (row, column) convention. The resulting LLG equation is 

\begin{align}
\dot{\vec{m}}_{\eta, ij}\left|_0\right.  &= \omega_{\eta,ij} \hat{\vec{z}}\times\vec{m}_{\eta, ij} + \alpha_{\eta,ij} \vec{m}_{\eta, ij} \times \dot{\vec{m}}_{\eta, ij} \nonumber \\ &+ J^{S}_{\eta}\vec{m}_{\eta, ij} \times (\vec{m}_{\eta, ij} \times \hat{\vec{z}}). 
\label{Eqn:SingleDynamics}
\end{align}

The STO ferromagnetic resonance frequency is given by $\omega_{\eta,ij} = \gamma_{\eta,ij}(H_0-4\pi M_{\eta,ij})$ where $\gamma_{\eta,ij}$ is the gyromagnetic ratio and $M_{\eta,ij}$ is the saturation magnetization; $\alpha_{\eta,ij} \ll 1 $ is the dimensionless Gilbert damping parameter. We assume the resonance frequency and the Gilbert damping to the be same for all STOs in the array and drop the subscripts going forward, i.e. $\omega_{\eta,ij} \rightarrow \omega$ and $\alpha_{\eta,ij} \rightarrow \alpha$. The third term in Eq.~\eqref{Eqn:SingleDynamics} is the Slonczewski-Berger spin-transfer torque~\cite{berger1996, slonczewski1996}. The injected spin current $J^{S}_{\eta}$ is assumed to the be same for all sites in a given sublattice. 

Interactions between adjacent STOs are mediated by a reactive Ruderman- Kittel-Kasuya-Yosida (RKKY)-type magnetic exchange coupling, which is given by~\cite{Flebus2020} 

\begin{align}
\dot{\vec{m}}_{A, ij}\left|_{\rm coup}\right. &= -\vec{m}_{A, ij} \times (J\vec{m}_{B, ij} + \tilde{J}\vec{m}_{B,ij-1}) \nonumber \\ &-\vec{m}_{A, ij} \times J_2(\vec{m}_{\eta', i+1j} + \vec{m}_{\eta',i-1j})
\end{align}

\begin{align}
\dot{\vec{m}}_{B, ij}\left|_{\rm coup}\right. = -\vec{m}_{B, ij} \times (J\vec{m}_{A, ij} + \tilde{J}\vec{m}_{A,ij+1}) \nonumber \\ -\vec{m}_{B, ij} \times J_2(\vec{m}_{\eta'', i+1j} + \vec{m}_{\eta'',i-1j})
\end{align}

\begin{figure}[t!]
\begin{center}
\includegraphics[width=3in]{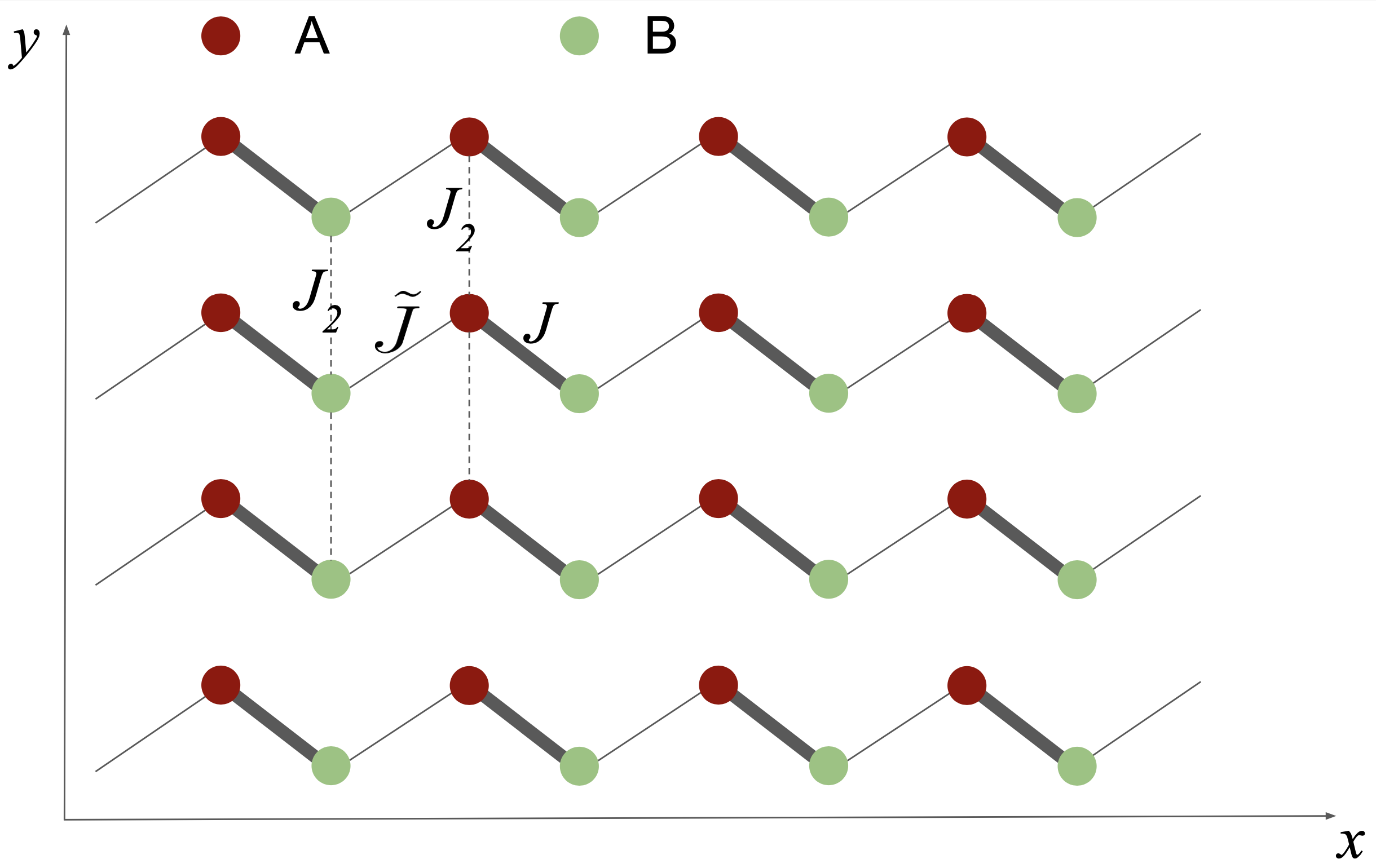}
\end{center}
\caption{Schematic of the effective tight-binding model of a 2d STO array, showing a system of size $4\times 8$, where there are $4$ unit cells per row. Red (dark color) sites are the A sublattice and green (light color) sites are the B lattice.}
\label{fig:schematic}
\end{figure}

The coupling strengths $J, \tilde{J}$ are the intracell and intercell coupling for the 1d unit cell, which contains two STOs, and $J, \tilde{J}, J_2 > 0 $ indicate ferromagnetic coupling. We consider the geometry depicted in Fig.~\ref{fig:schematic}, for which $\eta' = A, \eta'' = B$.

We now derive the linearized equations of motion for the STO magnetization $\vec{m}_{A(B)}$. That is, we start in the strong-field regime where the magnetic moment is mostly aligned along the $\hat{\vec{z}}$ direction and the energy scale given by $\omega$ is the largest one in the problem. Starting from Eq.~\eqref{Eqn:SingleDynamics}, we can derive an effective Hamiltonian in the following way: First, we linearize the equations of motion about the equilibrium field direction by writing $\vec{m}_{\eta, ij} = (m^x_{\eta, ij}, m^y_{\eta, ij}, 1)^T$ where $|m^x_{\eta, ij}| \simeq |m^y_{\eta, ij}| \ll 1$, and $|m^z| \simeq 1$ is assumed to be a constant. We then introduce the variable $2m^-_{\eta, ij} = m^x_{\eta, ij} - im^y_{\eta, ij}$, and make the Holstein-Primakoff approximation $m^-_{A(B), ij} = \langle a_{ij}(b_{ij})\rangle e^{-i \omega t}$ where the second quantized bosonic operators $a_{ij}$, $b_{ij}$ annihilate a magnon on the STO at site $(i,j)$~\cite{Auerbach2012}. From the Heisenberg equations of motion $\dot{a}_{ij}  = i/\hbar\left[H, a_{ij}\right]$, we derive an effective Hamiltonian corresponding to the linearized equations of motion. In the following we set $\hbar = 1$. 

\begin{figure*}[ht]
\centering
\includegraphics[width=\textwidth, 
                     keepaspectratio]{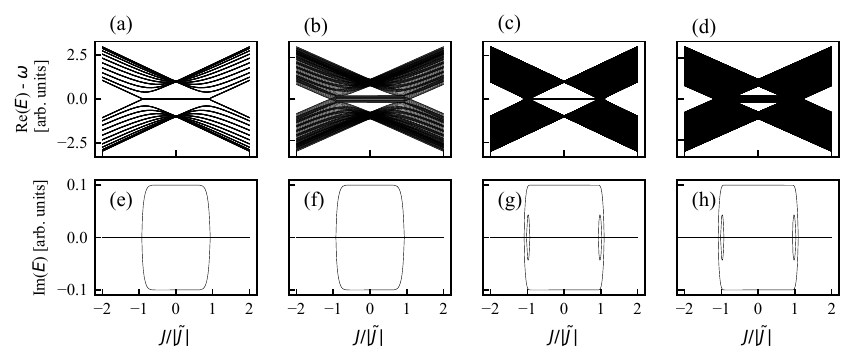}
    \caption{The dependence of the real (a-b) and imaginary (e-f) parts of the energy spectrum of $H$ on $J/|\tilde{J}|$ for a system of  $10 \times 20$ STOs for $\omega = 1\tilde{J}$, $\alpha = 0.2$, and $J_2 = 0.01\tilde{J}$~(a,e) and $J_2 = 0.1\tilde{J}$~(b,f). Modes with $\mathrm{Im}(E) > 0$ correspond to auto-oscillation of STOs on the edge of the system, appearing for $|J| < \tilde J$. These modes correspond to the flat in-gap bands at $\mathrm{Re}(E)-\omega \approx 0$ in the real spectrum. Panels (c-d, g-h) show the real (c-d) and imaginary (g-h) parts of the energy spectrum for a system of $50 \times 100$ STOs, with $J_2 = 0.01\tilde{J}$~(c,g) and $J_2 = 0.1\tilde{J}$~(d,h). Apart from the different system size, all the other parameters are same as in (a-b) and (e-f). In the larger system there is some activation of bulk STOs in the \PT-broken regime, indicated in (g-h) by the additional states with $\mathrm{Im}(E)> 0$ in the region $\alpha\omega > |\tilde{J} - J|$.\label{fig:spectrum1}} 
\end{figure*}

For the 2d STO array, $H = \sum_{ij} H_{ij}$ is
\begin{align}
H_{ij}&=\omega  (a^\dagger_{ij}a_{ij}+ b^\dagger_{ij}b_{ij})\nonumber\\
&+i (J^{S}_{A}-\alpha\omega) a^\dagger_{ij}a_{ij}+i (J^{S}_{B}-\alpha\omega) b^\dagger_{ij}b_{ij}\nonumber\\
&-J(a^\dagger_{ij}b_{ij}+h.c.)-\tilde J( a^\dagger_{ij}b_{ij-1}+h.c.)\nonumber\\
&-J_2(a^\dagger_{ij}a_{i-1j}+b^\dagger_{ij}b_{i-1j}+h.c). \label{eqn:real_spaceH}
\end{align}

We see that non-Hermiticity arises due to the onsite spin current injection and Gilbert damping, resulting in onsite terms $\propto i(J^{S}_\eta - \alpha\omega)$, and the degree of non-Hermiticity can be tuned by balancing the injected spin current and Gilbert damping. Given the non-Hermitian lattice model, we can now explore its energy spectrum and edge state properties.  In this manuscript we use the term `energy spectrum' and the symbol $E$ to denote the complex eigenvalues of $H$; Re($E$) can be thought of as an energy while $\mathrm{Im}(E) > 0$ ($< 0$) is an indication of lasing (damping).  

Here we consider the effect of different coupling strengths $J_2$. We confine the system to the parity-time (\PT) symmetric regime where the injected spin current is  $J_{SA} = 2\alpha\omega$, $J_{SB} = 0$; further symmetry analysis is performed in Sec.~\ref{sec:symmetry}. This is a 2d extension of the 1d \PT-symmetric, non-Hermitian SSH model analyzed in Ref.~\cite{Flebus2020}. We note that other versions of the non-Hermitian 2d SSH model studied in the literature have alternating A and B sublattice sites in the vertical direction~\cite{Obana2019, liu2019topologically, yuce2019topological}, resulting in a four site unit cell. Here, however, we introduce the coupling such that each column is all A or B sublattice sites, thus reducing the unit cell to two STOs. Our reasoning for this geometry is that it would be easier to inject spin current into an entire column of STOs, for example using a metallic strip, rather than having to individually address each A or B site STO in the grid.  

\section{Results \label{Sec:Results}}

\begin{figure}[t!]
\begin{center}
\includegraphics[width=0.5\textwidth]{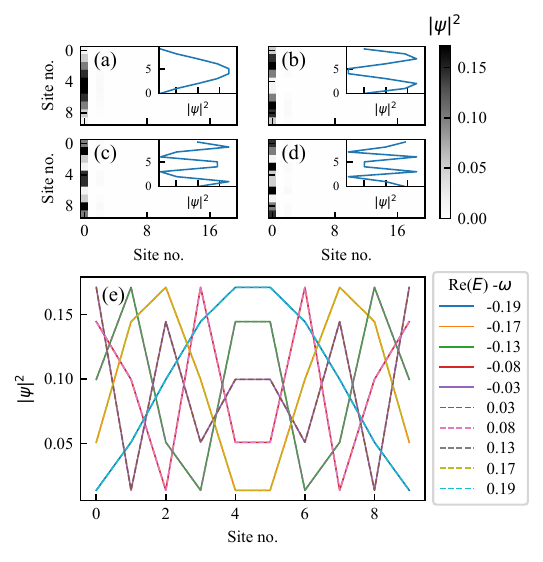}
\end{center}
\caption{Spatial distribution of 1d edge modes in a 2d STO array. (a-d) Color density plots showing $|\psi|^2$ for four of the the flat-band modes with $\mathrm{Im}(E) > 0$. The states are localized on the left edge, with $|\psi|^2 = 0$ everywhere in the bulk. The insets show the 1d distribution along the edge. (e) 1d plot showing the spatial distribution of all 10 lasing edge states and the corresponding real part of the energy, $\mathrm{Re}(E)-\omega$. The edge states are not uniform across the system due to the coupling $J_2$ between rows. States appear in pairs where states with equal magnitude and opposite sign of $\mathrm{Re}(E)-\omega$ have the same spatial distribution (dashed and solid lines overlap).}
\label{fig:10edgestates}
\end{figure}

Our goal is to understand how the vertical coupling $J_2$ between rows affects the properties of the original 1d system, thus we briefly review the results for a 1d STO chain~\cite{Flebus2020, Gunnink2022}. The real energy spectrum of the 1d STO model Hamiltonian has two degenerate flat bands for $J<|\tilde{J}|$ and admits a real line gap in $k$-space~\cite{kawabata2019}. The flat bands with degenerate energy eigenvalues are an indication of topologically protected edge states of the Hamiltonian. The eigenstates corresponding to the flat-bands also have non-zero imaginary eigenvalues, indicating lasing ($\mathrm{Im}(E) > 0$) and damped ($\mathrm{Im}(E) < 0$) states. The lasing states correspond to an auto oscillation of STOs, which occurs only at the edge of the system for $\alpha\omega < |J-\tilde{J}|$. In the regime $\alpha\omega > |J-\tilde{J}|$, the bulk Hamiltonian also has complex eigenvalues which can lead to oscillation of the bulk STOs; this is the so-called \PT-broken regime of the model where the Hamiltonian respects \PT-symmetry but its eigenstates do not~\cite{Hurst2022}. Symmetry analysis of the Bloch Hamiltonian for the 1d model confirms the auto-oscillation of the edge STOs to be a zero-dimensional `edge state' with topological protection. 

In the 2d case, the non-Hermitian Hamiltonian Eq.~\eqref{eqn:real_spaceH} also has complex eigenvalues. The energy spectrum of the 2d model looks similar to the 1d model in the case of weak vertical coupling $J_2/\tilde{J} \lesssim 0.01$, but the degeneracy in the flat bands breaks immediately even with infinitesimal vertical coupling. In Fig.~\ref{fig:spectrum1}, we show the energy spectrum of the Hamiltonian considering different vertical coupling strengths $J_2/\tilde{J} = 0.01, 0.1$ for two different system sizes. We found that as $J_2$ increases with respect to $\tilde{J}$, the energy separation between flat bands increases and they hybridize with the bulk states. We simulated results for a 2d array 20 sites wide and 10 sites in the vertical direction, as well as a larger system 100 sites wide and 50 sites in the vertical direction. In the upcoming subsections, we respectively discuss the numerical results for the edge states, analyze the Hamiltonian in momentum space, and present a symmetry analysis of the model.

\subsection{Edge States} 
Using exact diagonalization, we find that the real-space Hamiltonian Eq.~\eqref{eqn:real_spaceH} exhibits one-dimensional `lasing' edge states where $\mathrm{Im}(E) > 0$. In Fig.~\ref{fig:10edgestates} we show the spatial distribution of edge states for a system of $10\times 20$ STOs. Since there are 10 rows, the system exhibits 10 lasing edge modes, all with $\mathrm{Im}(E) = 0.09$. We have confirmed that the corresponding damped edge modes with $\mathrm{Im}(E) = -0.09$ occur on the opposite edge of the system, as expected (not pictured here). We set the intercell coupling to $\tilde{J} = 1$ and consider the regime where intracell coupling $J/\tilde{J}= 0.2$ and vertical coupling $J_2/\tilde{J}=0.1$. The physical manifestation of these edge states is an auto-oscillation of STOs that is non-uniform at the edge of the sample. Thus, the STOs will exhibit spatially varying microwave emission which can be tuned based on the magnon population in each edge mode. 
\*
\begin{figure}[t!]
\begin{center}
\includegraphics[width=0.5\textwidth]{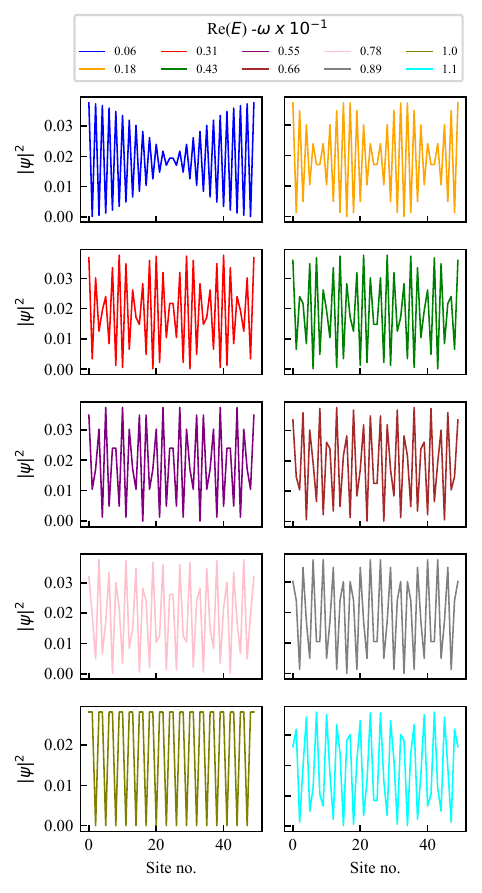}
\end{center}
\caption{Example of the spatial distribution of 10 unique lasing edge states (out of 50) and the corresponding real part of the energy, $\mathrm{Re}(E)-\omega \times 10^{-1}$ for a system of $50 \times 100$ sites. The larger system shows greater heterogeneity in the spatial distribution of edge modes. The top row of the legend corresponds to the left column and the bottom row corresponds to the right column; values increase downward in each column. For example $\mathrm{Re}(E)-\omega = 1.0$ corresponds to the bottom-left panel.}
\label{fig:50by100_left}
\end{figure}
\*
We also diagonalized a larger STO array to see how the results vary if we scale up the system size. The Hamiltonian matrix has a block banded structure, called a block Toeplitz-tridiagonal (TT) matrix. For a larger system i.e. 50x100 sites, the matrix becomes 5000 x 5000 for which the diagonalization is computationally expensive for different values of $J_2$.  To diagonalize this large sparse matrix efficiently, we used the method given by Ref.~\cite{ABDERRAMANMARRERO2020125324}. The energy spectrum for the larger system - as shown in Fig.~\ref{fig:spectrum1}~(c-d), (g-h) - doesn't show any deviation apart from very small bulk oscillations in the region where $\alpha\omega > |J - \tilde{J}|$. This proves the 1d edge states can exist in a significantly larger system. As shown in Fig.~\ref{fig:50by100_left}, the edge states in the larger system also exhibit a non-uniform spatial distribution, with increased oscillations and beating behavior visible in some modes. Here, we show the first 10 out of 50 total edge states ranked by increasing Re($E$). 

\subsection{Bloch Hamiltonian}
Here we consider the  \PT-symmetric regime where $J_{SA} = 2\alpha\omega$ and $J_{SB} = 0$, leading to balanced gain (loss) terms $\pm i\alpha\omega$ on the A(B) sites. To analyze the Hamiltonian in momentum space, we consider periodic boundary conditions and Fourier transform Eq.~\eqref{eqn:real_spaceH}, which is written as 
\begin{equation}
\hat H = \sum_{\vec{k}}
\begin{pmatrix} a^\dagger_{\vec{k}} &  b^\dagger_{\vec{k}} \end{pmatrix} 
\left(H_{\vec{k}} -\omega \right) 
\begin{pmatrix}  a_{\vec{k}} \\  b_{\vec{k}} \end{pmatrix} 
\end{equation}
where 
\begin{equation} 
H_{\vec{k}}=\left(\begin{array}{cc}
i\alpha\omega-2J_2\cos{k_y} & -J-\tilde J e^{ik_x} \\  -J-\tilde J e^{-ik_x} & -i\alpha\omega-2J_2\cos{k_y} \end{array}\right)
\label{Eqn:BlochH}
\end{equation} 
The resonant frequency simply provides an overall shift of the energy spectrum, therefore we redefine $\omega$ as the zero energy point.

The Bloch Hamiltonian can be written 
\begin{equation}
    H_{\vec{k}} = d_0(k_y)\mathbb{1} + \vec{d}(k_x)\cdot\boldsymbol{\sigma} \label{eqn:Bloch}
\end{equation}
where $\mathbb{1}$ is the $2\times 2$ identity matrix and $\boldsymbol{\sigma} = (\sigma^x, \sigma^y, \sigma^z)$ is the vector of Pauli matrices. We define the functions $d_0(k_y) = -2J_2\cos(k_y)$ and \begin{equation}
    \vec{d}(k_x) = \begin{pmatrix}
        -J-\tilde{J}\cos(k_x) \\ \tilde{J}\sin(k_x) \\ i \alpha \omega 
    \end{pmatrix}.
\end{equation}
We find the two-band energy spectrum 
\begin{equation}
    \epsilon_\pm(\vec{k}) = - 2J_2\cos(k_y) \pm \sqrt{J^2+\tilde{J}^2 + 2J\tilde{J}\cos(k_x) - \alpha^2\omega^2}.
    \label{Eq:spectrum-k}
\end{equation}

Here we see that the eigenvalues are real for $\alpha\omega < |J+\tilde J e^{ik_x}|$, i.e. the system remains in the \PT-unbroken regime exhibiting real eigenvalues as long as the Gilbert damping is relatively small. This condition is satisfied for all $k$ if $\alpha\omega < |J-\tilde{J}|$. Furthermore, the spectrum is linear in the vertical coupling $J_2$; thus the effect of coupling adjacent 1d STO chains together is to shift the spectrum away from the resonant frequency $\omega$. We can see this clearly for example in Fig.~\ref{fig:spectrum1}~(b), where the flat bands of adjacent 1d chains hybridize and are vertically shifted. Furthermore, $J_2$ can cause the energy gap to close, however in the regime $J_2 \lesssim \sqrt{(J-\tilde{J})^2-\alpha^2\omega^2}$ the energy gap is open and the edge states remain well separated from the bulk.

\subsection{Symmetry Analysis\label{sec:symmetry}}

Symmetry analysis can help determine whether the 1d edge states displayed in Figs.~\ref{fig:10edgestates} and~\ref{fig:50by100_left} are topologically protected. We investigate the following symmetries of the Bloch Hamiltonian, Eq.~\eqref{Eqn:BlochH}: chiral symmetry, chiral-inversion symmetry, sublattice symmetry, and parity-time (\PT) symmetry. Systems obeying chiral, chiral-inversion, or sublattice symmetry can exhibit topologically protected edge modes~\cite{Gong2018, jin2019bulk, kawabata2019}, therefore it is important to check whether these symmetries are preserved for our model. \PT-symmetry ensures there is always a regime in which the Hamiltonian has real eigenvalues~\cite{bender1999}. We note that for a Hermitian system, chiral and sublattice symmetries are equivalent, however for a non-Hermitian Hamiltonian this is no longer the case and some care must be taken. Here we use the symmetry naming conventions from Ref.~\cite{kawabata2019}. 

To have chiral symmetry (CS), the Hamiltonian must satisfy the condition $\sigma_z H_{\vec{k}}^\dagger\sigma_z=-H_{\vec k}$. 
We find that the vertical coupling term $J_2\cos(k_y)$ breaks chiral symmetry in general, however for the special values $k_y = \pm\pi/2$ chiral symmetry is preserved. To have chiral-inversion (CI) symmetry, the Hamiltonian must satisfy the condition $\sigma_yH_{\vec k}\sigma_y = -H_{-\vec{k}}$. Like the case of CS, CI is in general broken by $J_2$ and only preserved for $k_y = \pm\pi/2$. For sublattice symmetry, the Hamiltonian must satisfy the condition $\sigma_z H_{\vec{k}}\sigma_z=-H_{\vec k}$.  This model does not posses sublattice symmetry for any parameter regime due to the non-Hermitian terms, as is the case in 1d~\cite{Lieu2018, Flebus2020}. 

To have \PT-symmetry, the Hamiltonian must satisfy the condition $\sigma_xH_\vec{k}^*\sigma_x=H_{\vec k}$. This condition is satisfied for $J_{SA}=2\alpha\omega$ and $J_{SB}=0$, thus as with the 1d case the system can be tuned to the \PT-symmetric regime by altering the injected spin current on A and B sublattice sites. \PT-symmetry alone does not guarantee topological protection~\cite{Lieu2018}, and the results from the symmetry analysis indicate that the edge states observed in this model are not topologically protected. However, the confinement of the oscillatory modes to the edge can be understood as a result of \PT-symmetry in the bulk, which is broken spontaneously by the edge of the system. Furthermore, \PT-symmetry guarantees that the bulk STOs do not have any lasing modes as long as $\alpha\omega < |J-\tilde{J}|$.

\section{Conclusion and Outlook \label{Sec:Conclusion}} 

In this work we have examined a novel realization of a non-Hermitian 2d SSH model which can be constructed from an array of STOs. Using exact diagonalization and analysis of the Bloch Hamiltonian, we have shown that this model exhibits 1d lasing edge states with a non-uniform spatial distribution. The physical manifestation of these modes is an auto-oscillation of STOs along one edge of the system which is spatially varying. The extension of the model from 1d to 2d via the addition of vertical coupling between individual 1d STO chains breaks chiral-inversion, chiral, and sublattice symmetry, indicating a loss of topological protection for these modes. However, the vertical coupling preserves \PT-symmetry for the bulk states, thereby guaranteeing that the bulk oscillators do not activate, even in the presence of spin current injected into the bulk. 

Here we have considered an injected spin current such that the system remains at the \PT-symmetric point. Future works could investigate the robustness of these edge states in the presence of additional terms such as dipolar interactions as well as dissipative coupling between STOs. However, results from studies of the analogous 1d model indicate that if these terms are small compared to the reactive RKKY coupling studied here, they would not strongly affect the presence of edge states~\cite{Flebus2020, Gunnink2022}. 

Another interesting extension of this model would be to explore possible application in devices. The dispersion relation in Eq.~\eqref{Eq:spectrum-k} indicates that the 1d edge states have nonzero group velocity due to the coupling $J_2$.  Future work can examine whether these edge states, despite being non-uniform, could be used as a 1d channel for spin transport. For example, if there is additional spin injected into one of the edge STOs exhibiting lasing, one could study how it travels along the edge of the system. Such a device could potentially provide a new way to realize a low-dissipation transport channel for spin.

\begin{acknowledgments}
 H.M.H. acknowledges support of the San Jos\'{e} State University Research, Scholarship, and Creative Activity assigned time program. B.F. acknowledges support of the National Science Foundation under Grant No. NSF DMR-2144086.
\end{acknowledgments}

\bibliography{main}

\end{document}